\documentclass[aps,prd,titlepage,nofootinbib,preprint,superscriptaddress]{revtex4-2}

\usepackage[T1]{fontenc}
\usepackage[utf8]{inputenc}
\usepackage{lmodern}
\usepackage{amsmath}
\usepackage{amssymb}
\usepackage{mathtools}
\usepackage{graphicx}
\usepackage{xcolor}

\usepackage[hidelinks]{hyperref}

\begin{document}

\title{Stationary branches of non-canonical spinor theories and their cosmology}

\author{Ugo Moschella}
%\email{ugo.moschella@uninsubria.it}
\affiliation{Dipartimento di Scienza e Alta Tecnologia, Universit\`a degli Studi dell'Insubria, via Valleggio 11, I-22100 Como, Italy}
\affiliation{INFN, Sezione di Milano, via Celoria 16, I-20133 Milano, Italy}

\author{Mario Novello}
\affiliation{Centro de Estudos Avan\c{c}ados de Cosmologia (CEAC/CBPF), Rua Dr. Xavier Sigaud 150, CEP 22290-180, Rio de Janeiro, Brazil}

\author{Oliver F. Piattella}
\email{of.piattella@uninsubria.it}
\affiliation{Dipartimento di Scienza e Alta Tecnologia, Universit\`a degli Studi dell'Insubria, via Valleggio 11, I-22100 Como, Italy}
\affiliation{INFN, Sezione di Milano, via Celoria 16, I-20133 Milano, Italy}

\begin{abstract}
We study a classical spinor field minimally coupled to gravity and described by a non-canonical Lagrangian depending on the fermionic kinetic scalar and the scalar bilinear. For purely kinetic theories, we show that the field equations split the solutions into two distinct
branches. On the regular branch, the spinor satisfies the massless Dirac equation and contributes a vacuum energy determined by an additive constant in the Lagrangian. On the stationary branch, the spinor field equations become identically satisfied, while the energy-momentum tensor is exactly that of a cosmological constant for any spinor configuration and in any spacetime. The spinor evolution is
therefore not uniquely determined by the field equations. In a spatially flat homogeneous universe, both branches act as
cosmological-constant sources. We also give the dilution law for homogeneous configurations in theories with explicit dependence on the scalar bilinear. When a potential with a nonzero minimum is introduced, the stationary configuration becomes the late-time state of the cosmological evolution, and the approach to it generates a
pressureless component in addition to an effective cosmological constant. For a simple quadratic model, the background expansion reproduces exactly that of the standard $\Lambda$CDM cosmology, while the value of the vacuum energy remains controlled by an additive constant in the Lagrangian.
\end{abstract}

%\keywords{spinor cosmology, dark energy, non-canonical fields} % not printed by revtex4-2 without the showkeys option

\maketitle

\section{Introduction}\label{Sec:Introduction}

The observed accelerated expansion of the Universe is commonly attributed to dark energy \cite{SupernovaSearchTeam:1998fmf, SupernovaCosmologyProject:1998vns}. The cosmological constant is its simplest realization, but its small observed value remains theoretically unexplained \cite{Weinberg:1988cp}. This has motivated the study of additional matter components, including fields with non-canonical kinetic terms. For scalar fields, the best-known examples are $k$-inflation and $k$-essence \cite{Armendariz-Picon:1999hyi,Armendariz-Picon:2000ah,Armendariz-Picon:2000nfh}. In purely kinetic $k$-essence, a stationary point of the Lagrangian is approached dynamically, and the approach can describe a unified dark sector behaving as a cosmological constant plus pressureless matter \cite{Scherrer:2004au}.

Classical spinor fields have long been considered as cosmological sources \cite{Novello:1976ft, Armendariz-Picon:2003wfx, Ribas:2005vr, Saha:2011ya, Saha:2016cbu}. Their self-interactions can generate accelerated phases and nonsingular backgrounds, and spinor-driven bounces and their perturbations continue to be investigated \cite{Magueijo:2012ug, Farnsworth:2017wzr, Tukhashvili:2023itb, Fekecs:2025exl}. Non-canonical spinor theories with a Lagrangian depending nonlinearly on the fermionic kinetic scalar have been studied under the name of $f$-essence and $g$-essence \cite{Myrzakulov:2010fessence, Yerzhanov:2010gessence, Jamil:2011fessence, Inagaki:2011jk, Tsyba:2011ss}.

Our aim is to complete the classification of solutions of purely kinetic spinor theories, for an arbitrary metric and without imposing any symmetry on the spinor configuration. In the $g$-essence literature, the purely kinetic constraint was considered for homogeneous configurations, and only the root $X=0$ was retained \cite{Yerzhanov:2010gessence}. For a Lagrangian $L(X,S)$, where $X$ is the symmetrized fermionic kinetic scalar and $S=\bar\psi\psi$, the field equations imply the covariant identity $XL_X=-SL_S$, which, for purely kinetic theories, reduces to the constraint $XL_X=0$. Solutions on the regular branch have $X=0$ and $L_X(0)\neq0$ and obey the massless Dirac equation, whereas those on the stationary branch have $X=X_e$ with $L_X(X_e)=0$. On the stationary branch, the stress tensor is exactly that of a cosmological constant for any configuration and in any spacetime, but the spinor evolution is not determined by the field equations. On the regular branch, the vacuum energy is simply the additive constant $-L(0)$, while a stationary branch can support a positive vacuum energy even when $L(0)=0$. In a spatially flat Friedmann--Lema\^itre--Robertson--Walker (FLRW) spacetime, homogeneous configurations of either branch behave as a cosmological constant.

Finally, when a potential $V(S)$ with a minimum at $S_*\neq0$ is added to the purely kinetic Lagrangian, the stationary point becomes a late-time state of the cosmological evolution, and the approach produces a pressureless component in the spirit of Ref.~\cite{Scherrer:2004au}. For a quadratic model with a suitable relation among its parameters, the background is exactly that of $\Lambda$CDM, although the value of the vacuum energy is again fixed by an additive constant.

The paper is organized as follows. Section~\ref{Sec:Fermkessence} presents the theory and its energy-momentum tensor. In Sec.~\ref{Sec:purelykinetic}, we classify the solutions of purely kinetic theories in an arbitrary spacetime. Section~\ref{Sec:cosmology} specializes the equations to a spatially flat FLRW spacetime, and Sec.~\ref{Sec:homogeneouskinetic} discusses the homogeneous purely kinetic solutions. Theories with a potential $V(S)$ are studied in Sec.~\ref{Sec:deformed}. We summarize our conclusions and discuss open questions in Sec.~\ref{Sec:Conclusions}. The geometric conventions are collected in Appendix \ref{App:FLRWcalc}.

\section{Non-canonical spinor theory}\label{Sec:Fermkessence}

We consider the action
\begin{equation}\label{action}
I=\int d^4x\sqrt{-g}\left[\frac{R}{2\kappa}+L(X,S)\right]\; ,
\qquad \kappa=8\pi G\; ,
\end{equation}
where
\begin{equation}\label{definitions}
S\equiv\bar\psi\psi\; ,
\qquad
X\equiv\frac{i}{2}
\left[
\bar\psi\Gamma^\mu D_\mu\psi
-(D_\mu\bar\psi)\Gamma^\mu\psi
\right]\;,
\end{equation}
are the scalar bilinear and the symmetrized fermionic kinetic scalar, respectively, and $\psi$ is a commuting classical spinor. We work in units $c=\hbar=1$ with signature $(+---)$, and adopt the torsion-free (Levi-Civita) connection with the geometric conventions of Appendix~\ref{App:FLRWcalc}. The curved-space gamma matrices $\Gamma^\mu=e_a{}^\mu\gamma^a$ satisfy $\{\Gamma^\mu,\Gamma^\nu\}=2g^{\mu\nu}$, with $\bar\psi=\psi^\dagger\gamma^0$ and $\gamma^5=i\gamma^0\gamma^1\gamma^2\gamma^3$, and the spinor covariant derivatives are
\begin{equation}
D_\mu\psi=\partial_\mu\psi-\Omega_\mu\psi\; ,
\qquad
D_\mu\bar\psi=\partial_\mu\bar\psi+\bar\psi\Omega_\mu\; ,
\end{equation}
with
\begin{equation}
\Omega_\mu=\frac{1}{8}\omega_\mu{}^{ab}[\gamma_a,\gamma_b]\; .
\end{equation}
We refer to Ref.~\cite{Shapiro:2016pfm} for a detailed discussion of spinors in curved spacetime.

Independent variations of the action with respect to $\bar\psi$ and $\psi$ give
\begin{align}
iL_X\Gamma^\mu D_\mu\psi
+\frac{i}{2}(\nabla_\mu L_X)\Gamma^\mu\psi
+L_S\psi&=0\; ,
\label{psieqgen}\\
iL_XD_\mu\bar\psi\Gamma^\mu
+\frac{i}{2}(\nabla_\mu L_X)\bar\psi\Gamma^\mu
-L_S\bar\psi&=0\; .
\label{barpsieqgen}
\end{align}
Here and below, subscripts on $L$ denote its partial derivatives. 

% In particular,
% \begin{equation}\label{derivativeLX}
% \nabla_\mu L_X = L_{XX}\nabla_\mu X+L_{XS}\nabla_\mu S\; .
% \end{equation}
% Because $X$ already contains first derivatives of $\psi$, Eqs. \eqref{psieqgen} and \eqref{barpsieqgen} contain second derivatives through $\nabla_\mu X$ whenever $L_{XX}\neq0$.

Contracting Eq.~\eqref{psieqgen} from the left with $\bar\psi$ and Eq.~\eqref{barpsieqgen} from the right with $\psi$, and subtracting the results, yields the on-shell identity
\begin{equation}\label{onshellidentity}
XL_X = -SL_S\; .
\end{equation}
This is an algebraic relation between the two invariants $X$ and $S$, which defines a set of curves in the $(X,S)$ plane. Wherever $L_X+XL_{XX}+SL_{XS}\neq0$, these can be written locally as $X=X(S)$. In that case $\nabla_\mu X=(dX/dS)\nabla_\mu S$, and the second derivatives drop out of the field equations on shell. 

The vector current $J^\mu=\bar\psi\Gamma^\mu\psi$ is not conserved in general. Instead, the sum of the same two contractions gives
\begin{equation}\label{dressedcurrent}
 \nabla_\mu(L_XJ^\mu)=0\;,
\end{equation}
which is the conservation law of the Noether current associated with global phase invariance.

% On any region where $L_X\neq0$, one may define
% \begin{equation}
% \widetilde\psi=\sqrt{|L_X|}\,\psi\; .
% \end{equation}
% Equation~\eqref{psieqgen} then becomes
% \begin{equation}\label{tildeeom}
% i\Gamma^\mu D_\mu\widetilde\psi
% +\frac{L_S}{L_X}\widetilde\psi=0\; .
% \end{equation}
% Because $L_X$ depends on derivatives of $\psi$, this rescaling is not automatically an invertible change of field variables. It is singular at $L_X=0$ and cannot be used on the stationary branches
% discussed below.

\subsection{Energy-momentum tensor}

We define the energy-momentum tensor by $T_{\mu\nu}=(2/\sqrt{-g})\,\delta I_m/\delta g^{\mu\nu}$, where $I_m$ is the matter part of the action. With the curvature convention of Appendix~\ref{App:FLRWcalc}, the gravitational equations following from Eq.~\eqref{action} are
\begin{equation}\label{Einsteinequation}
R_{\mu\nu}-\frac12 Rg_{\mu\nu}=-\kappa T_{\mu\nu}\; ,
\end{equation}
and the torsion-free tetrad variation of the matter action gives
\begin{equation}\label{Tmunufermion}
T_{\mu\nu}
=L_X\mathcal K_{\mu\nu}-g_{\mu\nu}L\; ,
\end{equation}
where
\begin{equation}\label{Kmunudef}
\mathcal K_{\mu\nu}
\equiv\frac{i}{2}
\left[
\bar\psi\Gamma_{(\mu}D_{\nu)}\psi
-D_{(\mu}\bar\psi\Gamma_{\nu)}\psi
\right]\; ,
\end{equation}
and parentheses denote symmetrization with weight $1/2$. Since $g^{\mu\nu}\mathcal K_{\mu\nu}=X$, Eqs.~\eqref{onshellidentity} and \eqref{Tmunufermion} give the on-shell trace
\begin{equation}\label{traceT}
T^\mu{}_\mu=XL_X-4L=-SL_S-4L\; .
\end{equation}
Moreover, $\nabla^\mu T_{\mu\nu}=0$ on shell, as a consequence of the diffeomorphism invariance of the matter action.

Unlike scalar $k$-essence, where the gradient of the field defines a fluid four-velocity, a generic spinor configuration is not a perfect fluid. A natural frame is provided instead by the vector current. For a commuting Dirac spinor, the Fierz identities give \cite{Lounesto:2001}
\begin{equation}\label{Fierzmain}
J_\mu J^\mu=S^2+P^2\; ,
\qquad
J_\mu A^\mu=0\; ,
\qquad
A_\mu A^\mu=-J_\mu J^\mu\; ,
\end{equation}
where $A^\mu=\bar\psi\Gamma^\mu\gamma^5\psi$ is the axial current and $P=i\bar\psi\gamma^5\psi$. Whenever $S^2+P^2>0$, $J^\mu$ is therefore timelike, while $A^\mu$ is spacelike and orthogonal to it. For such configurations, heat flux and anisotropic stress relative to $J^\mu$ need not vanish away from a homogeneous background, and a $1+1+2$ decomposition adapted to $J^\mu$ and $A^\mu$ can be more informative than a perfect-fluid analogy \cite{Fabbri:2025ffi, Vignolo:2025lrs, Vignolo:2026lrs}.

A dependence of $L$ on the pseudoscalar $i\bar\psi\gamma^5\psi$ could be included, but it would not affect the purely kinetic results and we omit it for simplicity.

\section{Purely kinetic theories in a general spacetime}\label{Sec:purelykinetic}

We now specialize to purely kinetic theories,
\begin{equation}
L=L(X)\; ,
\end{equation}
without restricting the metric or the spinor configuration. The on-shell identity then reduces to
\begin{equation}\label{pureconstraint}
XL_X=0\; ,
\end{equation}
and Eq.~\eqref{traceT} gives $T^\mu{}_\mu=-4L(X)$. At any spacetime point, a solution must therefore satisfy either $X=0$ or $X=X_e$, where $X_e$ denotes any stationary point of $L$, $L_X(X_e)=0$. We assume that $L$ is smooth and that the roots of $XL_X=0$ are isolated. Continuity then forces $X$ to be constant on each connected spacetime region, so every solution belongs to one of the two classes discussed below, according to whether $L_X$ is nonzero or vanishes at the root. In particular, $X$ cannot evolve towards a stationary point, as it does in purely kinetic $k$-essence \cite{Scherrer:2004au}.

\subsection{Regular branch: \texorpdfstring{$X=0$}{X=0}}

Suppose first that
\begin{equation}
X=0\; ,
\qquad
0<|L_X(0)|<\infty\; .
\end{equation}
Since $L_X=L_X(0)$ is constant, the spinor equation reduces to
\begin{equation}\label{regularcovarianteom}
i\Gamma^\mu D_\mu\psi=0\; ,
\end{equation}
and the energy-momentum tensor is
\begin{equation}\label{regularcovariantstress}
T_{\mu\nu}=L_X(0)\mathcal K_{\mu\nu}-L(0)g_{\mu\nu}\; .
\end{equation}
Conversely, any solution of the massless Dirac equation has $X=0$ and solves the nonlinear spinor equations on this branch. On such solutions, $\mathcal K_{\mu\nu}$ is the energy-momentum tensor of a massless Dirac field. It is traceless, because $g^{\mu\nu}\mathcal K_{\mu\nu}=X=0$, and covariantly conserved, because $\nabla^\mu T_{\mu\nu}=0$ on shell and $L(0)$ and $L_X(0)$ are constants. The regular branch thus describes a massless Dirac field, whose stress tensor is weighted by $L_X(0)$, together with the vacuum energy density $\rho_0=-L(0)$. For $L_X(0)>0$ and coupled to gravity, it is the Einstein--Dirac system with cosmological constant $\kappa\rho_0$, while for $L_X(0)<0$ the Dirac stress tensor enters with the opposite sign. In both cases, the spinor evolution is determined by initial data.

The stress tensor reduces to that of a cosmological constant only when $\mathcal K_{\mu\nu}=0$. This holds for the homogeneous configurations of Sec.~\ref{Sec:homogeneouskinetic}, but in general the spinor adds a traceless, radiation-like component.

\subsection{Stationary branch: \texorpdfstring{$L_X(X_e)=0$}{LX(Xe)=0}}

Consider an isolated stationary point
\begin{equation}
X=X_e\; ,
\qquad
L_X(X_e)=0\; .
\end{equation}
Because $X=X_e$ everywhere, $L_X$ vanishes identically and so does $\nabla_\mu L_X$. For a purely kinetic theory, both field equations then vanish identically. The only condition on $\psi$ is $X[\psi,g]=X_e$, a single real equation for the eight real components of $\psi$. Nevertheless,
\begin{equation}\label{stationarystress}
T_{\mu\nu}=-L(X_e)g_{\mu\nu}\; ,
\qquad
\rho_e=-L(X_e)\; ,
\qquad
p_e=L(X_e)\; ,
\end{equation}
for any configuration on the branch and any metric. In the absence of other matter, Eq.~\eqref{Einsteinequation} then reduces to
\begin{equation}\label{Einsteinspace}
R_{\mu\nu}=\kappa\rho_e\,g_{\mu\nu}\; .
\end{equation}
The metric is therefore an Einstein space with cosmological constant $\kappa\rho_e$, for example de Sitter or Schwarzschild--de Sitter spacetime when $\rho_e>0$. Conversely, we now show that any spacetime admits, at least locally, spinor configurations with $X=X_e$.

Under a local phase transformation $\psi\to e^{-i\theta}\psi$, with $\theta$ real, $J^\mu$ is invariant and
\begin{equation}\label{phaseshift}
X\to X+J^\mu\partial_\mu\theta\; .
\end{equation}
Let $\chi$ be any nowhere-vanishing spinor field on a given spacetime. Its current $J^\mu[\chi]$ is causal and future-directed, since $J_\mu J^\mu\geq0$ by Eq.~\eqref{Fierzmain} and its time component in any tetrad is $\chi^\dagger\chi>0$, and it is therefore transverse to any spacelike hypersurface. Hence the linear equation
\begin{equation}\label{transport}
J^\mu[\chi]\,\partial_\mu\theta=X_e-X[\chi]
\end{equation}
can be integrated, at least locally, along the integral curves of $J^\mu[\chi]$ once $\theta$ is prescribed on a spacelike hypersurface $\Sigma$. The spinor $\psi=e^{-i\theta}\chi$ then lies on the stationary branch. The branch is thus nonempty on any spacetime, and the spinor is not fixed by the field equations. Since $\chi$ is arbitrary, two solutions can coincide in a neighborhood of $\Sigma$ and differ elsewhere, while the metric obeys Eq.~\eqref{Einsteinspace} in both cases. Initial data on $\Sigma$ therefore do not determine the spinor evolution, in any spacetime. Real rescalings $\psi\to f\psi$, under which $X\to f^2X$ and $S\to f^2S$, provide a similar freedom whenever $X[\chi]X_e>0$.

For later use, we note that the construction extends to Lagrangians depending on $S$. Let $(X_c,S_c)$, with $S_c\neq0$, be a critical point of $L(X,S)$, where $L_X=L_S=0$. Configurations with $X=X_c$ and $S=S_c$ everywhere have $\nabla_\mu L_X=0$, so Eqs.~\eqref{psieqgen} and \eqref{barpsieqgen} vanish identically, and $T_{\mu\nu}=-L(X_c,S_c)g_{\mu\nu}$. Because $S$ is invariant under the phase transformation, such configurations can be built from any $\chi$ with $S[\chi]S_c>0$. One first applies the rescaling $f=\sqrt{S_c/S[\chi]}$ and then a phase that solves Eq.~\eqref{transport}, with $\chi$ replaced by $f\chi$ and $X_e$ by $X_c$. For the deformed theories of Sec.~\ref{Sec:deformed}, $(X_e,S_*)$ is such a critical point.

\section{Homogeneous spinor cosmology}\label{Sec:cosmology}

We now consider the spatially flat FLRW metric
\begin{equation}
ds^2=dt^2-a(t)^2\delta_{ij}dx^idx^j\; ,
\end{equation}
and a homogeneous spinor $\psi=\psi(t)$. The tetrad specified in Appendix~\ref{App:FLRWcalc} gives
\begin{equation}
\Omega_0=0\; ,
\qquad
\Omega_i=\frac{1}{2}\dot a\,\gamma_0\gamma_i\; ,
\qquad
X=\frac{i}{2}
\left(
\bar\psi\gamma^0\dot\psi
-\dot{\bar\psi}\gamma^0\psi
\right)\;,
\end{equation}
and the field equations become
\begin{align}
iL_X\gamma^0\dot\psi
+\frac{3iH}{2}L_X\gamma^0\psi
+\frac{i}{2}\dot L_X\gamma^0\psi
+L_S\psi&=0\; ,
\label{FLRWpsi}\\
iL_X\dot{\bar\psi}\gamma^0
+\frac{3iH}{2}L_X\bar\psi\gamma^0
+\frac{i}{2}\dot L_X\bar\psi\gamma^0
-L_S\bar\psi&=0\; .
\label{FLRWbarpsi}
\end{align}
The homogeneous energy-momentum tensor has the perfect-fluid background form
\begin{equation}
T^\mu{}_\nu=\operatorname{diag}(\rho,-p,-p,-p)\; ,
\end{equation}
with
\begin{equation}\label{fermionicrhoandp}
\rho=XL_X-L\; ,
\qquad
p=L\; .
\end{equation}
Indeed, for a homogeneous spinor the spin-connection terms cancel in the symmetrized $\mathcal K_{ij}$, so that $\mathcal K_{ij}=0$, while $\mathcal K_{00}=X$. For the mixed components, Eqs.~\eqref{FLRWpsi} and \eqref{FLRWbarpsi} give
\begin{equation}
L_X\mathcal K_{0i}
=-\frac{L_S}{4}\bar\psi\{\Gamma_i,\gamma^0\}\psi=0\;,
\end{equation}
so that $T_{0i}=L_X\mathcal K_{0i}$ vanishes on any solution, although $\mathcal K_{0i}$ itself need not vanish when $L_X=0$.

The Friedmann equations are therefore
\begin{equation}
3H^2=\kappa\rho\; ,
\qquad
2\dot H+3H^2=-\kappa p\;,
\end{equation}
with $H=\dot a/a$. The two spinor equations also imply directly
\begin{equation}\label{Sconservationdifferential}
L_X\dot S+(\dot L_X+3HL_X)S=0\; ,
\end{equation}
or, equivalently,
\begin{equation}\label{Sconservation}
a^3L_XS=C_S\; ,
\end{equation}
where $C_S$ is constant. In the same way, Eq.~\eqref{dressedcurrent} gives $a^3L_X\psi^\dagger\psi=\mathrm{constant}$. This is the specialization to $L(X,S)$ of the dilution law previously obtained in the $f$-essence literature \cite{Myrzakulov:2010fessence}. Together with Eq.~\eqref{onshellidentity}, Eq.~\eqref{Sconservationdifferential} implies the continuity equation $\dot\rho+3H(\rho+p)=0$, as required by the Friedmann equations. 

Note that Eq.~\eqref{Sconservationdifferential} was derived without dividing by $L_X$. On a solution for which $L_X$ vanishes identically, Eq.~\eqref{Sconservation} remains true with $C_S = 0$, but it gives no information about $S$.

% Using Eq.~\eqref{onshellidentity}, the equation-of-state parameter can be written as
% \begin{equation}\label{wgeneral}
% w=\frac{p}{\rho}
% =-\frac{1}{1+SL_S/L}\;.
% \end{equation}

\subsection{Canonical kinetic term and a potential}

For
\begin{equation}
L=X-V(S)\; ,
\end{equation}
Eqs.~\eqref{onshellidentity}, \eqref{fermionicrhoandp}, and \eqref{Sconservation} give the well-known relations \cite{Armendariz-Picon:2003wfx, Ribas:2005vr}
\begin{equation}
X=SV_S\; ,
\qquad
\rho=V\; ,
\qquad
p=SV_S-V\; ,
\qquad
S=\frac{S_0}{a^3}\; .
\end{equation}
More explicitly, the spinor solution is
\begin{equation}
\psi(t)=a(t)^{-3/2}
\exp\left[-i\gamma^0\int^t V_S(t')\,dt'\right]\psi_0\; ,
\end{equation}
where $\psi_0$ is a constant spinor and $S_0=\bar\psi_0\psi_0$.
The mass potential $V=mS$ reproduces pressureless matter with positive energy density when $mS>0$. The potential
\begin{equation}
V(S) = \left(A+BS^{1+\alpha}\right)^{1/(1+\alpha)}\; ,
\end{equation}
gives
\begin{equation}
w=-\frac{A}{A+BS^{1+\alpha}}\; ,
\end{equation}
and hence the generalized Chaplygin gas background, after using $S\propto a^{-3}$ \cite{Kamenshchik:2001cp, Bento:2002ps, Tsyba:2011ss}. For the usual interpolation from a matterlike regime to accelerated expansion, one may take $S>0$, $A,B>0$, and $\alpha>-1$.

\section{Homogeneous purely kinetic solutions}\label{Sec:homogeneouskinetic}

We now apply the classification of Sec.~\ref{Sec:purelykinetic} to a homogeneous spinor in the FLRW background of Sec.~\ref{Sec:cosmology}. On the regular branch, Eq.~\eqref{regularcovarianteom} becomes
\begin{equation}
\dot\psi+\frac{3H}{2}\psi=0\; ,
\end{equation}
and therefore
\begin{equation}\label{regularpsisolution}
\psi=\frac{\psi_0}{a^{3/2}}\; .
\end{equation}
Since $\mathcal K_{00}=X=0$, $\mathcal K_{ij}=0$, and $\mathcal K_{0i}=0$, the energy-momentum tensor is
\begin{equation}
T_{\mu\nu}=-L(0)g_{\mu\nu}\; ,
\qquad
\rho_0=-L(0)\; ,
\qquad
p_0=L(0)\; .
\end{equation}
A positive vacuum energy requires $L(0)<0$. It is entirely determined by the additive constant already present in the Lagrangian $L(X)$. The nonzero solution \eqref{regularpsisolution} is a stealth background in the sense that its amplitude does not affect $\rho_0$ or $p_0$.

On the stationary branch with $\rho_e>0$ and no other matter, Eq.~\eqref{Einsteinspace} gives de Sitter expansion with $3H^2=\kappa\rho_e$, and the freedom described in Sec.~\ref{Sec:purelykinetic} is already visible for homogeneous spinors. For $\psi(t)=\mathcal{A}(t)e^{-i\theta(t)}\xi$, with $\xi$ a constant spinor normalized by $\xi^\dagger\xi=1$, one finds $X=\mathcal{A}^2\dot\theta$. This is Eq.~\eqref{phaseshift} with $\chi=\mathcal{A}\xi$, for which $X[\chi]=0$ and $J^0[\chi]=\mathcal{A}^2$. Every smooth amplitude $\mathcal{A}(t)>0$, together with $\dot\theta=X_e/\mathcal{A}^2$, therefore provides a solution with the same energy-momentum tensor. Different choices of $\mathcal{A}(t)$ give different densities $J^0=\mathcal{A}^2$, so the spinor evolution is not fixed by the field equations, even when the metric is.

If one imposes $L(0)=0$, the regular branch carries no vacuum energy, and a cosmological constant can arise only on a stationary branch, which requires a nonlinear $L(X)$. For example, consider the Lagrangian
\begin{equation}\label{exampleL}
L(X)=X+\frac{X^2}{2M^4}\; ,
\end{equation}
where $M>0$ is a mass scale. The stationary point is
\begin{equation}
X_e=-M^4\;,
\end{equation}
and on this branch,
\begin{equation}
L(X_e)=-\frac{M^4}{2}\; ,
\qquad
\rho_e=\frac{M^4}{2}\; ,
\qquad
p_e=-\frac{M^4}{2}\; .
\end{equation}
The same Lagrangian also admits the regular branch, with $\rho_0=0$. Since $X$ is constant along any solution, a configuration cannot pass from one branch to the other.

The condition $L(0)=0$ is a choice of the additive constant, not a vacuum-energy selection mechanism: replacing $L$ by $L+C$ leaves the stationary roots unchanged and shifts $\rho_e$ by $-C$. No protection against such a shift is provided by the present classical model.

At the background level, the stationary construction resembles the mechanism of scalar ghost condensation \cite{ArkaniHamed:2003uy} and the late-time state of purely kinetic $k$-essence \cite{Scherrer:2004au}. There is, however, an essential difference. For a homogeneous scalar field $\phi$ with Lagrangian $P(Y)$ and $Y=\dot\phi^2/2$, the field equation gives $a^3P_Y\sqrt{Y}=\mathrm{constant}$, so that $P_Y\propto a^{-3}Y^{-1/2}$, and the stationary point is approached dynamically. In the spinor case, the relation \eqref{pureconstraint} is algebraic and locks $X$ to a root, so no such approach is possible within $L(X)$. We show in Sec.~\ref{Sec:deformed} that an approach of this kind is recovered once $L$ also depends on $S$.

\section{Deformed theories and the approach to the stationary point}\label{Sec:deformed}

The stationary point can be reached dynamically once the Lagrangian depends also on $S$. We consider
\begin{equation}\label{deformedL}
L(X,S)=F(X)-V(S)\; ,
\end{equation}
where $F$ has a nonzero stationary point $X_e$, with $F_X(X_e)=0$ and $F_{XX}(X_e)\neq0$, and $V$ has a minimum at $S_*\neq0$, with $V_S(S_*)=0$ and $V_{SS}(S_*)>0$. Wherever $F_X\neq0$, the homogeneous background is governed by the algebraic system
\begin{equation}\label{deformedsystem}
XF_X=SV_S\; ,
\qquad
a^3F_XS=C_S\; ,
\end{equation}
with
\begin{equation}
\rho=XF_X-F+V\; ,
\qquad
p=F-V\; .
\end{equation}
For $C_S\neq0$, $L_X\neq0$ at any finite $a$, and the spinor evolution is determined by the initial spinor once a branch of the system \eqref{deformedsystem} is chosen, in contrast with the stationary branch.

As $a\to\infty$, the product $F_XS$ must vanish. Consider a branch of the curve $XF_X=SV_S$ that ends at $(X_e,S_*)$, and write $X=X_e+\delta$ and $S=S_*+\eta$. To leading order, Eqs.~\eqref{deformedsystem} give
\begin{equation}
F_{XX}(X_e)S_*\,\delta=\frac{C_S}{a^3}\; ,
\qquad
X_eF_{XX}(X_e)\,\delta=S_*V_{SS}(S_*)\,\eta\; ,
\end{equation}
with corrections of order $a^{-6}$. Consequently,
\begin{equation}\label{latetimeexpansion}
\rho=\rho_\Lambda+\frac{\rho_\mathrm{m0}}{a^3}+O(a^{-6})\; ,
\qquad
p=-\rho_\Lambda+O(a^{-6})\; ,
\end{equation}
where
\begin{equation}
\rho_\Lambda\equiv V(S_*)-F(X_e)\; ,
\qquad
\rho_\mathrm{m0}\equiv\frac{X_eC_S}{S_*}\; .
\end{equation}
On such a branch, the stationary point is the late-time state of the homogeneous evolution, and its approach contributes a pressureless component, which is positive when $C_S$ has the sign of $X_e/S_*$. This is the spinor counterpart of the mechanism of Ref.~\cite{Scherrer:2004au}. 

Note that the condition $S_*\neq0$ is essential. For $V=\mu S^n$ with $n>0$, the branch ends at $(X_e,0)$, and the same expansion gives $\delta\propto a^{-3n/(n+1)}$. The deviation from $\rho_\Lambda$ then behaves as a fluid with $w=-1/(n+1)$, and a mass term ($n=1$) gives $w=-1/2$.

At late times, $S\to S_*$ while $L_X=C_S/(a^3S)$ decays as $a^{-3}$, so the critical point $(X_e,S_*)$ is reached only asymptotically. Since $L_X$ multiplies the ordinary Dirac kinetic term, spinor fluctuations become increasingly strongly coupled at late times. Whether the pressureless component clusters like cold dark matter requires a perturbative analysis that is beyond the scope of this work.

\begin{figure}
\centering
\includegraphics[width=\textwidth]{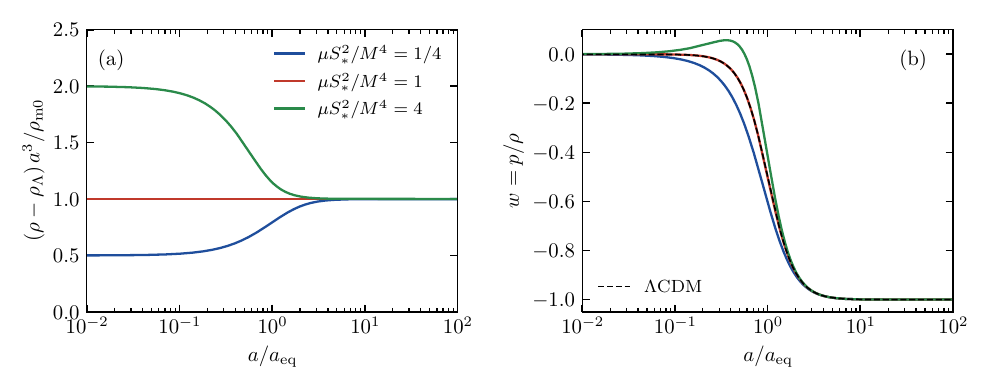}
\caption{Homogeneous background of the deformed model with $F$ given by Eq.~\eqref{exampleL} and $V$ given by Eq.~\eqref{quadraticV} with $V_0=0$, for the three values $\mu S_*^2/M^4 = 1/4,1,4$. (a) Effective pressureless density $(\rho-\rho_\Lambda)a^3$, normalized to its late-time value $\rho_\mathrm{m0}=M^4|C_S|/S_*$. (b) Equation-of-state parameter. The dashed line is the $\Lambda$CDM background with the same $\rho_\Lambda=M^4/2$ and $\rho_\mathrm{m0}$, which coincides with the case $\mu S_*^2/M^4=1$. The scale factor is normalized to $a_\mathrm{eq}$, defined by $\rho_\mathrm{m0}a_\mathrm{eq}^{-3}=\rho_\Lambda$.}
\label{Fig:deformed}
\end{figure}

As an example, take $F$ as in Eq.~\eqref{exampleL} and
\begin{equation}\label{quadraticV}
V(S)=\frac{\mu}{2}\left(S-S_*\right)^2+V_0\; ,
\end{equation}
with $\mu>0$, $S_*>0$, and a constant $V_0$. Then $X_e=-M^4$, $\rho_\Lambda=M^4/2+V_0$, and $\rho_\mathrm{m0}=M^4|C_S|/S_*$ for $C_S<0$. The relevant branch has $X<-M^4$ and $S>S_*$, and along it $\rho+p=XF_X>0$. At early times, $|X|\to\infty$ and $\rho\simeq M^2\sqrt{\mu}\,|C_S|\,a^{-3}$ with $p/\rho\to0$. The background therefore interpolates between two pressureless regimes whose amplitudes differ by the factor $\sqrt{\mu}\,S_*/M^2$, with a cosmological constant emerging at late times. For
\begin{equation}
\mu=\frac{M^4}{S_*^2}\; ,
\end{equation}
the curve $XF_X=SV_S$ contains the straight line $S/S_*=X/X_e$, and along it
\begin{equation}\label{exactLCDM}
\rho=\rho_\Lambda+\frac{M^4|C_S|}{S_*a^3}\; ,
\qquad
p=-\rho_\Lambda\; ,
\end{equation}
exactly, for all $a$. With this relation among the parameters, the model thus reproduces the $\Lambda$CDM background exactly. For other values of $\mu$, the effective pressureless amplitude interpolates between its early and late values, as shown in Fig.~\ref{Fig:deformed}, and for $\mu S_*^2>M^4$ the equation of state becomes positive during the transition, with a maximum $w\simeq0.06$ for $\mu S_*^2/M^4=4$ and stiff behavior approached for $\mu S_*^2\gg M^4$.

Two remarks are in order. First, $\rho_\Lambda$ is again fixed by additive constants. With the normalization $L(0,0)=0$, that is, $V_0=-\mu S_*^2/2$, one finds $\rho_\Lambda=(M^4-\mu S_*^2)/2$. This is positive for $\mu S_*^2<M^4$ and vanishes for the exact $\Lambda$CDM family, which then describes pure dust. The deformation therefore explains the approach to the stationary point, not the value of the vacuum energy. Second, $F_X<0$ along the relevant branch, so the Dirac term enters with the opposite sign relative to its normalization at $X=0$. For the commuting classical spinor considered here, the consequences of this sign for stability can only be assessed by the perturbative analysis mentioned above, and it would in any case matter for any quantization of the model. 

\section{Conclusions}\label{Sec:Conclusions}

We have studied a commuting classical spinor described by a non-canonical Lagrangian $L(X,S)$. The field equations imply the on-shell identity $XL_X=-SL_S$ and, for a homogeneous spinor in a flat FLRW spacetime, the dilution law $a^3L_XS=\mathrm{constant}$. The latter reproduces a result known in the $f$-essence literature and reduces to $S\propto a^{-3}$ for a canonical kinetic term.

Our main result is a covariant classification of the solutions of purely kinetic theories. When the roots of $XL_X=0$ are isolated, continuity fixes $X$ on each connected region of an arbitrary spacetime, so $X$ cannot evolve towards a stationary point. On the regular branch, with $X=0$ and $L_X(0)\neq0$, the spinor obeys the massless Dirac equation. The stress tensor is that of a massless Dirac field plus the vacuum energy density $-L(0)$, which originates from an explicit additive constant. On a stationary branch, with $L_X(X_e)=0$, the field equations vanish identically and the stress tensor is exactly $-L(X_e)g_{\mu\nu}$ for any configuration, so that, in the absence of other matter, the metric is an Einstein space. The spinor is subject only to the condition $X=X_e$, which any nowhere-vanishing spinor can be made to satisfy by a local phase. Its evolution is therefore not determined by initial data in any spacetime. A positive vacuum energy can arise on a stationary branch even when $L(0)=0$, as illustrated by Eq.~\eqref{exampleL}. In a flat FLRW spacetime, homogeneous configurations of both branches are exact cosmological-constant sources. On the regular branch, the spinor is a stealth field diluting as $a^{-3/2}$, while on the stationary branch an arbitrary amplitude function remains undetermined.

Within the purely kinetic theory, the stationary point cannot be approached dynamically. A potential $V(S)$ with a minimum at $S_*\neq0$ changes this. The stationary point becomes a late-time state of the homogeneous evolution, and its approach adds a pressureless component to the vacuum energy, in analogy with purely kinetic $k$-essence. For the quadratic model of Sec.~\ref{Sec:deformed}, the background interpolates between two pressureless regimes, and for $\mu=M^4/S_*^2$ it coincides exactly with that of $\Lambda$CDM, although the value of the vacuum energy is again set by an additive constant.

Several questions are left for future work. The first concerns the underdetermined spinor evolution on the stationary branch, which calls for a constraint analysis of the theory, including inhomogeneous configurations. The second concerns perturbations. On the stationary branch $L_X$ vanishes, and in the deformed theories it decays as $a^{-3}$, so the ordinary Dirac kinetic term is absent or suppressed, and spinor fluctuations are expected to be strongly coupled. Whether the pressureless component of the deformed models clusters like cold dark matter is a further open issue. These questions will be addressed elsewhere.

The stationary construction supplies an exact cosmological-constant source through a degenerate spinor sector. It does not select the value of the vacuum energy, and its spinor evolution is not fixed by the field equations. These results delimit the sense in which non-canonical spinors can provide an alternative realization of a cosmological constant and, once deformed, of a unified dark sector at the background level.

\appendix

% \section{Fierz identities}\label{App:Fierzid}

% For completeness, define the real bilinears
% \begin{equation}
% S=\bar\psi\psi\; ,
% \qquad
% P=i\bar\psi\gamma^5\psi\; ,
% \qquad
% J^\mu=\bar\psi\Gamma^\mu\psi\; ,
% \qquad
% A^\mu=\bar\psi\Gamma^\mu\gamma^5\psi\; .
% \end{equation}
% For a commuting Dirac spinor and signature $(+---)$, the algebraic Fierz identities give
% \begin{equation}
% J_\mu J^\mu=S^2+P^2\; ,
% \qquad
% A_\mu A^\mu=-J_\mu J^\mu\; ,
% \qquad
% J_\mu A^\mu=0\; .
% \end{equation}
% Thus $J^\mu$ is timelike for a regular spinor, while $A^\mu$ is spacelike and orthogonal to it.
% These identities follow directly from the completeness of the Dirac matrix basis; see, for example, Ref.~\cite{Lounesto:2001}.

\section{Geometric conventions and the FLRW spin connection}\label{App:FLRWcalc}

We use the Levi-Civita connection and define the curvature by
\begin{equation}\label{curvatureconvention}
R^\rho{}_{\sigma\mu\nu}
=\partial_\nu\Gamma^\rho_{\mu\sigma}
-\partial_\mu\Gamma^\rho_{\nu\sigma}
+\Gamma^\rho_{\nu\lambda}\Gamma^\lambda_{\mu\sigma}
-\Gamma^\rho_{\mu\lambda}\Gamma^\lambda_{\nu\sigma}\; ,
\qquad
R_{\sigma\nu}=R^\rho{}_{\sigma\rho\nu}\; .
\end{equation}
The tetrad connection appearing in $D_\mu\psi=\partial_\mu\psi-\Omega_\mu\psi$ is fixed by
\begin{equation}\label{tetradpostulate}
\partial_\mu e^a{}_\nu
-\Gamma^\lambda_{\mu\nu}e^a{}_\lambda
-\omega_\mu{}^a{}_b e^b{}_\nu=0\; .
\end{equation}
With $\Omega_\mu=\tfrac18\omega_\mu{}^{ab}[\gamma_a,\gamma_b]$, this convention gives the spinor derivatives used in the main text.

For
\begin{equation}
ds^2=dt^2-a(t)^2\delta_{ij}dx^idx^j\; ,
\end{equation}
we choose
\begin{equation}
e^{(0)}{}_\mu=\delta^0_\mu\; ,
\qquad
e^{(i)}{}_\mu=a\,\delta^i_\mu\; .
\end{equation}
The nonvanishing Christoffel symbols are
\begin{equation}
\Gamma^0_{ij}=a\dot a\,\delta_{ij}\; ,
\qquad
\Gamma^i_{0j}=H\delta^i_j\; .
\end{equation}
These conventions give
\begin{equation}
R_{00}=3(\dot H+H^2)\; ,
\qquad
R=6(\dot H+2H^2)\; ,
\qquad
R_{00}-\frac12R=-3H^2\; ,
\end{equation}
so Eq.~\eqref{Einsteinequation} yields $3H^2=\kappa\rho$ with the action and stress-tensor definition stated in Sec.~\ref{Sec:Fermkessence}. The corresponding spin connection is
\begin{equation}
\Omega_0=0\; ,
\qquad
\Omega_i=\frac{1}{2}\dot a\,\gamma_0\gamma_i\; .
\end{equation}
For a homogeneous spinor, the spatial spin-connection terms cancel in the symmetrized scalar $X$, leaving
\begin{equation}
X=\frac{i}{2}
\left(
\bar\psi\gamma^0\dot\psi
-\dot{\bar\psi}\gamma^0\psi
\right)\; .
\end{equation}
They nevertheless produce the $3H/2$ terms in Eqs.~\eqref{FLRWpsi} and \eqref{FLRWbarpsi}.

\bibliographystyle{apsrev4-2}
\bibliography{NLspinorcosmo-v9}

\end{document}